\documentclass[conference]{IEEEtran}
\IEEEoverridecommandlockouts
\usepackage{amsmath,amssymb,amsfonts}
\usepackage{algorithmic}
\usepackage{graphicx}
\usepackage{textcomp}
\usepackage{xcolor}
\usepackage{float}
\usepackage{tabularx,graphicx,url}
\usepackage{booktabs, longtable, placeins}
\usepackage{xcolor, graphics, multirow, multicol}
\graphicspath{{figures/}}
\usepackage{cancel}
\usepackage{siunitx}
\usepackage{threeparttable}
\usepackage[inline]{enumitem}
\def\BibTeX{{\rm B\kern-.05em{\sc i\kern-.025em b}\kern-.08em
    T\kern-.1667em\lower.7ex\hbox{E}\kern-.125emX}}

\begin{document}
\pagestyle{plain}
\title{How Does a Single EEG Channel Tell Us About Brain States in Brain-Computer Interfaces?\\
\thanks{Zaineb AJRA received a doctoral fellowship from \mbox{AXIAUM} Univ. Montpellier-ISDM ({\small ANR-20-THIA-0005-01}) and ED I2S in France.}
}

\author{\IEEEauthorblockN{Zaineb Ajra$^{1}$, Binbin Xu$^{2}$, Gérard Dray$^{2}$, Jacky Montmain$^{2}$, Stephane Perrey$^{1}$}
\IEEEauthorblockA{$^{1}$EuroMov Digital Health in Motion, Univ. Montpellier, IMT Mines Ales, Montpellier, France.}
\IEEEauthorblockA{$^{2}$EuroMov Digital Health in Motion, Univ. Montpellier, IMT Mines Ales, Ales, France.}
}

\maketitle

\begin{abstract}
Over recent decades, neuroimaging tools, particularly electroencephalography (EEG), have revolutionized our understanding of the brain and its functions. EEG is extensively used in traditional brain-computer interface (BCI) systems due to its low cost, non-invasiveness, and high temporal resolution. This makes it invaluable for identifying different brain states relevant to both medical and non-medical applications. Although this practice is widely recognized, current methods are mainly confined to lab or clinical environments because they rely on data from multiple EEG electrodes covering the entire head. Nonetheless, a significant advancement for these applications would be their adaptation for ``real-world" use, using portable devices with a single-channel. In this study, we tackle this challenge through two distinct strategies: the first approach involves training models with data from \textit{multiple channels} and then testing new trials on data from a \textit{single channel} individually. The second method focuses on training with data from a \textit{single channel} and then testing the performances of the models on data from all the other channels individually. To efficiently classify cognitive tasks from EEG data, we propose Convolutional Neural Networks (CNNs) with only a few parameters and fast learnable spectral-temporal features. We demonstrated the feasibility of these approaches on EEG data recorded during mental arithmetic and motor imagery tasks from three datasets. We achieved the highest accuracies of 100\%, 91.55\% and 73.45\% in binary and 3-class classification on specific channels across three datasets. This study can contribute to the development of single-channel BCI and provides a robust EEG biomarker for brain states classification.
\end{abstract}

\section{Introduction}

Brain-computer interface (BCI) provides a non-muscular communication pathway with external devices by decoding brain activities into computer control signals \cite{wolpaw2007brain}. BCI can be helpful in restoring some communication ability for patients with traumatic brain disorders or other specific neurological diseases \cite{luaute2015bci}. Neuroscientists have shown considerable attention to the evolution of BCIs, especially for patients with motor disabilities such as disorder of consciousness (DoC) patients. BCI-based technologies can be used to restore motor function as a tool for neuro-rehabilitation in clinical settings and as a means of communication with their environment \cite{fuchs2003neurofeedback}.

Electroencephalography (EEG) is one of the most common neuroimaging methods used in BCI research \cite{lotte2007review}. EEG allows the measurement of brain electrical activity without requiring necessarily a behavioral response from the subject, which could be helpful as a diagnostic tool, especially for patients with severe brain injury. Therefore, analyzing and interpreting EEG activity patterns provide an excellent way to study cognitive functions. It can also help researchers understand the neural mechanisms underlying human behavior and help people increase their productivity and well-being. Consequently, the first step is to better classify the acquired EEG signals, for example, by separating EEG signals containing cognitive activities from the resting state.

Recent advancements in BCI technology have highlighted the need for simpler, more portable EEG systems, particularly those using single-channel EEG for various applications including motor imagery classification, cognitive assessments, and sleep monitoring. Multi-channel EEG setups, although providing high spatial resolution, suffer from drawbacks such as high costs, complex configurations, and reduced mobility, making them impractical for daily use and routine clinical applications. Conversely, single-channel EEG offers a promising alternative due to its ease of use, reduced computational load, and potential for integration into wearable technologies. This approach not only simplifies the hardware requirements but also substantially lowers the barrier for non-technical users. Moreover, studies have demonstrated the feasibility of using single-channel EEG for effective feature extraction, even in the face of challenges such as the loss of spatial information typically obtained from multiple electrodes \cite{ozmen2018biologically}, \cite{kanoga2018comparative}, \cite{cretot2023assessing}. The development of robust decoding algorithms and feature extraction techniques specifically adapted to single-channel EEG is therefore critical. This focus aligns with the growing interest in deploying BCIs in real-world settings, where the balance between system simplicity and classification accuracy is crucial. The exploration of single-channel EEG configurations, supported by deep learning frameworks, presents a transformative potential for BCIs, promoting wider adoption and facilitating the development of low-cost, user-friendly systems that preserve reasonable accuracy and functionality for everyday applications.

Although motor imagery-based BCI (MI-BCI) is very useful in clinical applications for control and rehabilitation \cite{barsotti2015full}, non-motor tasks such as mental arithmetic (MA-BCI) are considered more appropriate than standard BCI paradigms because of their wide accessibility and less strict technical requirements.

Raw EEG signals are only one-dimensional, requiring further processing to obtain useful features. Typically, three techniques are used to evaluate the EEG signal: time-domain analysis, frequency-domain analysis, and time-frequency analysis.
For time-frequency-based analysis, a spectrogram image is another way to represent the features of raw EEG data. Different magnitudes in the spectrogram reflect varying energy values and the frequency responses over time. Spectrograms may have better performance in a classification task on signal/time-series data compared to other handcrafted feature extraction techniques, as they contain more unknown and valuable features \cite{yuan2018patients}.

In computer vision tasks such as image classification, deep convolutional neural networks have shown significant advantages \cite{o2020deep}. However, deeper neural networks and complex architectures require much larger datasets, resulting in heavy computational loads, which might not be suitable for applications with relatively small datasets. Shallow neural networks are thus the first step to explore the feasibility of related tasks.

The main objective of this work is to explore the application of single-channel EEG data in detecting mental activity to support wearable technology. While multi-channel EEG and various physiological data have shown reliability in detecting mental activity, the idea of using single-channel EEG data to determine complex neurological phenomena is relatively novel. Our pipeline is structured to enhance the classification performance of single-channel EEG, thus improving both the practicality and efficiency of wearable BCIs.
To address these challenges, this study employs two distinct approaches to EEG data analysis:
\begin{enumerate}
    \item  Training on all channels, then testing on single channels -- The CNN model is trained on data from all available EEG channels but is tested on each channel individually to determine its effectiveness in isolating relevant features.
    \item  Training on a single channel, then testing on single channels -- The CNN model is trained on data from an individual channel and tested across all other channels to assess their generalizability and robustness in diverse conditions.
\end{enumerate} 
These methodologies are designed to validate the feasibility of using single-channel EEG in practical settings and to refine the techniques for improved accuracy.

To the best of our knowledge, existing EEG-based BCI methodologies mainly involve training on all available channels and testing on individual ones to isolate relevant features \cite{van2018single}, or directly extracting features for classification from single-channel EEG, typically applied to specific tasks such as imagining hand movements or engaging in visual activities \cite{ko2017development}, \cite{ge2014classification}. However, training models on individual channels and testing them across others -- remains largely unexplored. This approach is designed to rigorously assess generalizability and robustness of the model across different settings, potentially enhancing the adaptability and reliability of EEG-based diagnostics in a wide range of applications.

\section{Materials and methods}

\subsection{Data from cognitive tasks}
In this study, three EEG databases were used, denoted as Dataset 1, Dataset 2, and Dataset 3 respectively.
\subsubsection{Dataset 1}  This dataset comes from \cite{shin2018eyes} and was studied in our previous work \cite{ajra2022mental}. The dataset comprises EEG recordings from 22 electrodes. 12 Participants engaged in a series of 60 trials, each consisting of a mental arithmetic (MA) task and a baseline (BL) period, across three sessions. The experimental protocol included visual instructions and alternating periods of task performance and rest, structured around a pre-rest and post-rest phase of 15 seconds each. 

\subsubsection{Dataset 2} 
The second dataset comes from \cite{Shin2017Open} consisting of EEG data from 29 healthy subjects. The dataset included 30 active electrodes. The experiments were conducted in three sessions of MA and BL tasks. Each session consisted of 20 repetitions of the task phase (27--\SI{29}{\second}). During each experiment, participants were instructed to look at a visual instruction displayed on the monitor during \SI{2}{\second} indicating the type of task, followed by \SI{10}{\second} of the task, and from \SI{15}{\second} to \SI{17}{\second} of resting period. The raw EEG signals from 30 electrodes were re-referenced using a common average reference, down-sampled to 200 Hz, and then band-pass filtered between 0.5-50 Hz. Ocular artifacts were removed using the automatic artifact rejection toolbox in EEGLab. The artifact-free data was then segmented into epochs using time intervals from \SI{-2}{\second} (\SI{2}{\second} before the stimulus) to \SI{10}{\second} (after the stimulus). In total, 60 epochs with a duration of \SI{12}{\second} each were obtained for each participant across the three sessions.

\subsubsection{Dataset 3}
The third dataset was collected by \cite{ma2020multi}. The MI-2 dataset consisting of EEG data from 25 right-handed healthy subjects. The dataset comprises three distinct classes: `rest', `hand' and `elbow'. Each 8-second trial began with a 2-second display of a white circle, followed by a red circle for 1 second to announce the upcoming `hand' or `elbow' stimulus, displayed for 4 seconds. Participants were invited to imagine the movement without physically moving, concluding with a `Break' screen for 1 second. Overall, the experiment comprised 19 sessions, 15 with 40 trials each for MI tasks, and 4 with 75 trials each for resting, resulting a total of 900 trials per subject (each category of MI has 300 trials).
The MI-2 dataset included 64 active electrodes. The raw EEG signals were band-pass filtered between 0.5-100 Hz and downsampled to 200 Hz. The data was spatially filtered using the Common Average Reference (CAR) method and time-domain filtered within the 0 to 40 Hz range. 

\begin{figure}[!htb]
   \centering
  \includegraphics[width=\columnwidth]{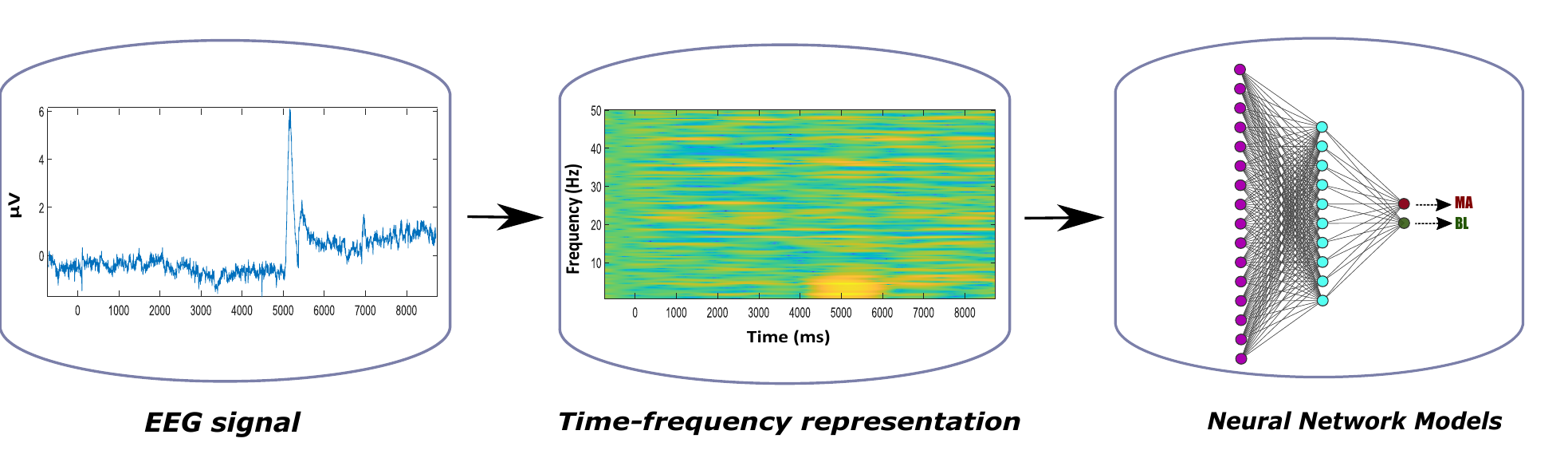}
   \caption{The workflow of the MA-EEG signal classification process in this study. From raw EEG signals to different Neural Network models
}
   \label{fig:workflow}
\end{figure}

\subsection{Spectral-temporal feature extraction}

The time-frequency analysis enables the extraction of richer features from EEG data \cite{
tsinalis2016automatic}. Spectrograms were generated from pre-processed EEG signals using the event-related spectral power (ERSP) approach with the EEGLab toolbox. ERSP was applied to each epoch, and the generated spectrogram was fixed in size at $224 \times 224$ to match the input size of common deep neural network models. In total, $15\,480$ and $52\,200$ spectrograms were generated from datasets 1 and 2, respectively. For dataset 3, we selected 25 (subjects) x 62 (channels) x 4 (sessions) x 5 (epochs) for each task. A total of $93\,000$ spectrograms were generated from dataset 3. The overall workflow of this study is illustrated in Figure \ref{fig:workflow}.

\subsection{Convolutional Neural Network and Classification}

A shrinkage linear discriminant analysis (shrinkage LDA) was used by \cite{shin2018eyes} and \cite{Shin2017Open} to distinguish a mental arithmetic and motor imagery tasks from a rest period. They extracted features using CSP filters applied to EEG data. On the other hand, MA et al. (2020) extracted the features using Filter Bank CSP (FBCSP)  and employed a Support Vectror Machine (SVM) as classifer to classify each subject's data into a three-class scenario.

In our study, we proposed a shallow convolutional neural network (CNN) for all datasets to maintain consistency in model application. For datasets specifically related to the mental arithmetic tasks from \cite{shin2018eyes} and \cite{Shin2017Open}, we expanded the use of our model to include basic recurrent neural network (RNN) and common state-of-the-art deep neural networks (GoogLeNet and ShuffleNet). Note that these models were trained them from scratch.

\subsubsection{Architecture of the Proposed CNN}

CNN architectures automatically carry out feature extraction and classification. When compared to manual feature extraction approaches, CNN performs quite well \cite{ozturk2020stacked}. Efficient model design depends on several factors, such as layers of CNN architectures, configurations, and the conditions required for the training step \cite{nour2021novel}. In the literature, 1D CNN architectures are often used for solving signal processing problems, 3D CNN architectures for dealing with volume image problems, and 2D CNN architectures are used in general to solve image-related tasks. The main advantage of 2D CNN models is that this architecture takes into account two-dimensional features from images (both frequency and temporal in the case of spectrogram), which may yield better features. 

\begin{figure}[!htb]
\centering
\includegraphics[width=0.6\columnwidth]{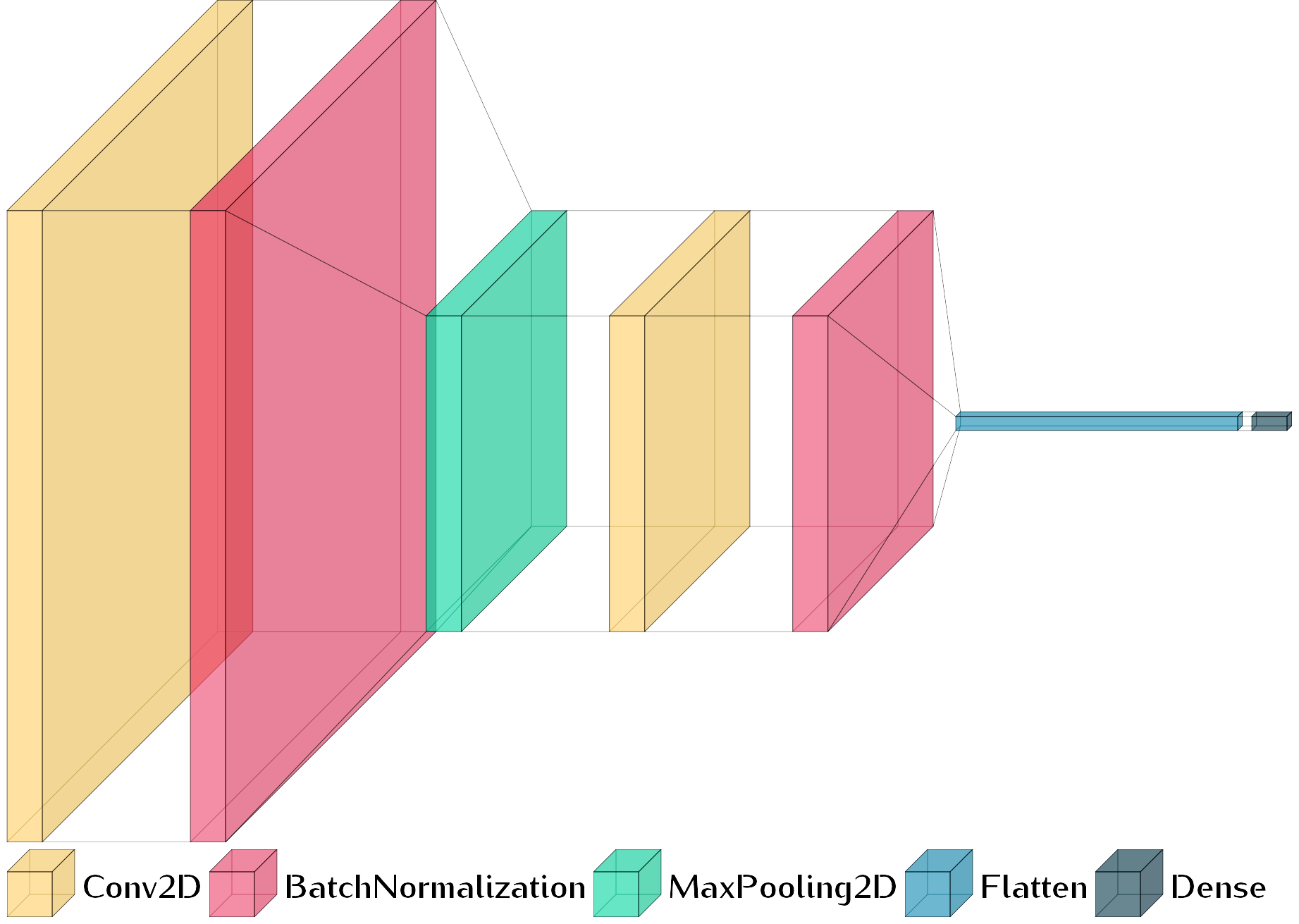}
\caption{CNN model architecture}
\label{fig:model_cnn}
\end{figure}

The proposed shallow CNN model (Figure \ref{fig:model_cnn}) is composed essentially of two blocks of 2D convolutional layers (Conv2D) to extract the most relevant features from images. Each convolutional layer is attached to a batch normalization layer to speed up training and to improve the convergence of the model. Both Conv2D layers use a Rectified Linear Unit (ReLU) activation and have 50 filters with a kernel size of $ 5 \times 5$ to disrupt the network's linear structure and make it sparse. These two blocks of Conv2D are linked by a max pooling layer with a size of $ 2 \times 2$ aiming to reduce the number of parameters in the network. 

\subsection{Training options and performance evaluation}

The proposed CNN model was trained with the following configuration: Stochastic gradient descent with momentum (SGDM) was used to optimize the model. The initial learning rate was fixed at 0.001. The maximum number of training epochs was 100, with a batch size of 64. In a previous study \cite{ajra2022mental}, the early stopping rule was strictly applied to avoid the issue of over-fitting. In the current study, the early stopping rule was adjusted according to the model architectures. 

The $15\,840 $, $52\,200$ and $92\,000$ trials were randomly split into {\sc training} (70\%), {\sc validation} (15\%), and {\sc test} (15\%) sets, taking into account their own specific subject, channel, and event (class) ratios. To minimize the risk of over-fitting, 10 random triple-sets were generated. The median values of commonly used metrics (Accuracy, Sensitivity, Specificity, and F1-score) were reported from this cross-validation. Only metrics from the {\sc test} set obtained from the final epoch are reported below. 

It is important to note that the same data-splitting and model training policy was applied in all three datasets. The analysis was performed in a more comprehensive way, with updated results.

\section{Results / Discussion}
\label{sec:results}

\subsection{Multi-channels training, single channel testing}

In a previous study, we demonstrated the robustness of the proposed shallow neural network, achieving a classification performance of 90.68\% in cross-validation. As shown in Table \ref{tab:model_perf_3_datatsets}, we observed that the classification performance could be further improved by applying less strict rules and with a larger number of cross-validation models (50 random triple-sets instead of 20 as in \cite{ajra2022mental}). With the refined pipeline, we were able to achieve a cross-validation accuracy of 93.75\% which significantly exceeded the baseline results obtained with shrinkage linear discriminant analysis (sLDA).

\begin{table}[!htb]
\setlength{\tabcolsep}{4pt}
  \centering
\caption{EEG performance (median $\%$) across datasets: updates compared to previous study \cite{ajra2022mental} and performance on datasets \cite{shin2018eyes}, \cite{Shin2017Open} and \cite{ma2020multi}} 
  \footnotesize
    \begin{tabular}{ccccccc}
    \toprule
    \textbf{Dataset }& \textbf{Methods } & \textbf{Task } & \textbf{ACC} & \textbf{Sensitivity} & \textbf{Specificity} & \textbf{F1 score} \\
    \midrule
    \cite{shin2018eyes}& sLDA & MA & 80.10& -- &  --  & -- \\
    \cite{Shin2017Open}& sLDA & MA & 75.90 & -- & --  & -- \\ 
    \cite{Shin2017Open}& sLDA & MI & 65.60 & --  & -- & --  \\
    \cite{ma2020multi} & SVM & MI & 68.68 & --  & -- & -- \\
    \midrule
    \cite{shin2018eyes}& CNN & MA & \textbf{93.75} & 93.94 &  93.43  & 93.73 \\
    \cite{Shin2017Open}& CNN & MA & \textbf{91.57} & 91.77 & 91.48  & 91.50 \\ 
    \cite{Shin2017Open}& CNN & MI & \textbf{87.80} &  88.02 & 87.59 & 87.83  \\
    \cite{ma2020multi}& CNN & MI & \textbf{86.75} & 86.92  & 93.14 & 86.25  \\
  \bottomrule
    \end{tabular}
  \label{tab:model_perf_3_datatsets}
\end{table}

The original data classifier proposed by \cite{Shin2017Open} used 10-fold cross-validation with sLDA for a binary classification problem. The sLDA model achieved average classification accuracies of 75.9\% and 65.6\%, for MA and MI tasks, respectively. However, the shallow CNN achieved a much better performance of classification, with median accuracies of 91.57\% for MA 87.80\% for MI tasks. 

The initial data classifier introduced by \cite{ma2020multi} implemented 5-fold cross-validation using an SVM to handle a three-class classification problem. The SVM model achieved an average accuracy of 68.68\% in classifying MI tasks. In contrast, the proposed CNN far outperformed it, recording a median accuracy of 86.75\% in the same three-class condition.

For datasets \cite{shin2018eyes} and \cite{Shin2017Open} related to mental arithmetic tasks, alternative deep learning models were evaluated. State-of-the-art models such as ShuffleNet and GoogleNet demonstrated similar or slightly improved performance, achieving accuracies of 91.87\% and 92.72\%, respectively. Despite their potential, the significant increase in computational load for a marginal improvement of 1\% makes these models less practical for monitoring applications. Moreover, the performance of LSTM was much less satisfactory, with a median accuracy of 72.99\%, showing no improvement even without early stopping, due to its inability to capture spatial correlations in the data.

\begin{figure}[!htb]
   \centering
  \includegraphics[width=\columnwidth]{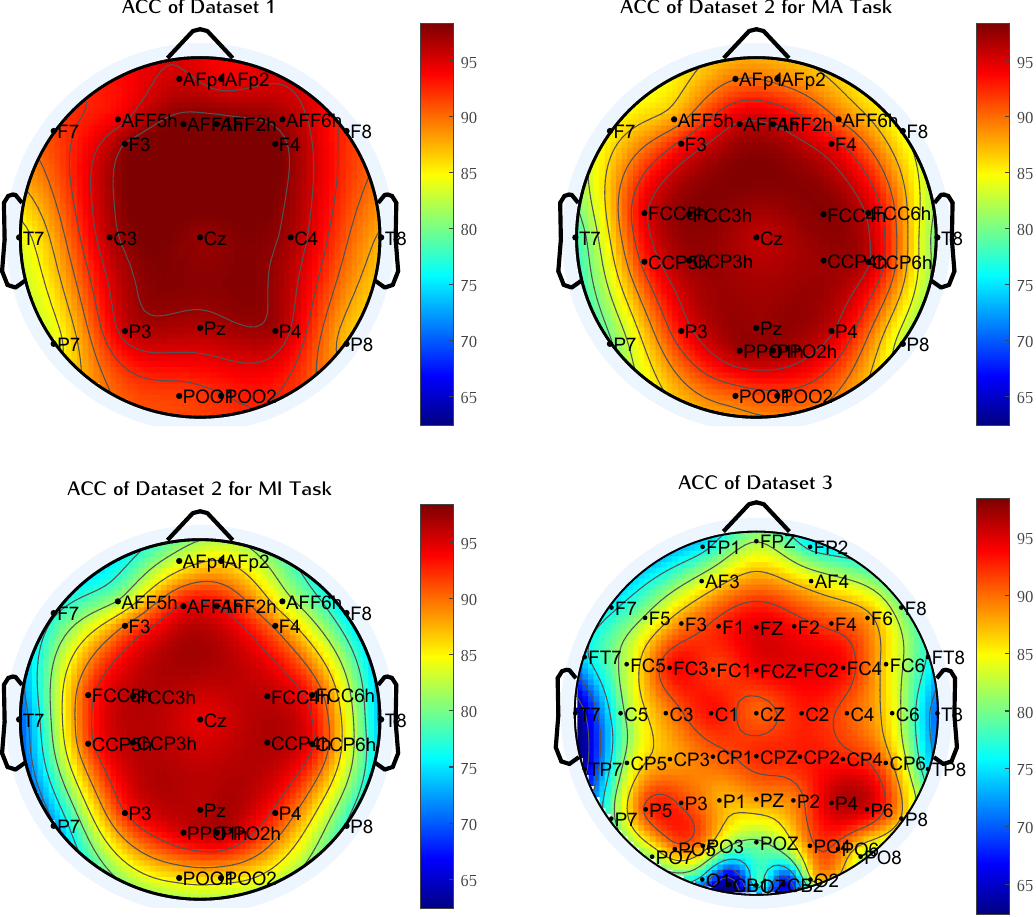}
   \caption{Topological representation of accuracy across channels for each dataset}
   \label{fig:topoplot}
\end{figure}

To optimize the computation time of the network and the number of electrodes while maintaining significant performance, many researchers have tried to reduce the number of electrodes used by classifying each individual channel based on its accuracy performance. In this study, we employed a strategy in which the CNN model is exhaustively trained on data from all available EEG channels to capture a broad spectrum of spatial patterns and correlations between channels. This training approach exploits the full potential of multi-channel data, enabling the model to discern and exploit the complex inter-channel dynamics characteristic of EEG recordings. The model is then tested on each channel individually, leveraging its ability to generalize in order to isolate and use the features most relevant for accurate classification. By evaluating the discrimination strength at the channel level, we can effectively rank channels according to their individual contributions to the model’s performance. The classification accuracy values for each channel are shown in Figure \ref{fig:topoplot}. 

\begin{table}[ht]
\setlength{\tabcolsep}{5pt}
\centering
\caption{Top and Low Test Channel Performance (median \%) by Trained Channel with CNN, training on single-channel, test on single-channel dataset \cite{shin2018eyes}}
\scalebox{0.9}{
\begin{tabular}{p{0.12\columnwidth}|p{0.25\columnwidth}|p{0.07\columnwidth}|p{0.35\columnwidth}|p{0.07\columnwidth}}
\toprule
\textbf{Trained Channel }& \textbf{Top predicted\newline Channel} & \textbf{Top \newline ACC} & \textbf{Low predicted channel \newline Channel} & \textbf{Low  ACC} \\ \midrule
F7 & AFp2  & 84.16  & F8 &  70.83 \\
AFF5h & AFp2  & 88.33 & P3 & 78.33 \\
F3  & AFp2 &  89.16 & AFF1h & 80.83 \\
AFp1 & Pz & 91.66 & F7, AFF6h &  85.83  \\ 
AFp2 & AFp2, P8 & 92.50 & T7 &  86.66   \\ 
AFF6h & F8 & 93.33 & F3 & 88.33  \\ 
F4 & AFF5h & 94.16 & AFF1h & 86.66  \\ 
F8 & AFF5h, AFp2, C3 & 95.00 & AFF1h & 90.83   \\ 
AFF1h & F4, F8, AFF2h, T7 & 95.83 & AFF6h & 91.66  \\ 
AFF2h & F4, AFF2h, Cz & 96.66 & AFF6h &  92.50   \\ 
Cz & F8, C3 & 97.50 & AFF6h & 92.50  \\       
Pz & AFp1, AFF2h & 98.33 & POO1 & 93.75 \\ 
T7 & AFp1, AFF2h & 98.33 & POO1 &  94.16 \\ 
C3 & AFp1, C3, P7, T8 & 98.33 & P4 & 95.41 \\ 
P7 & T8 & 99.16 & AFF5h & 95.83 \\ 
P3 & P7 & 99.16 & F4 & 95.83  \\ 
POO1 & P7 & 100 & F7 & 97.08  \\ 
POO2 & P7 & 100 & AFF5h, AFF1h, C4 & 97.50 \\ 
P4 & AFF6h, P7 & 100 & T8 & 97.50   \\ 
P8 & AFp1, F8, T7, P3 & 100 & AFF5h, AFp2, Cz, P4, C4, T8 & 98.33  \\ 
C4 & F3, AFF2h, C3, P4 & 100 & C4 & 97.91 \\ 
T8 & F3, AFp2, AFF6h, F4, AFF1h, AFF2h, P4 & 100 & F7, P8 & 98.33    \\ 
\bottomrule
\end{tabular}
}
\label{tab:perf_channel_approach2_shin2018}
\end{table}

\begin{table}[ht]
\setlength{\tabcolsep}{5pt}
\centering
\caption{Top and Low Test Channel Performance (median \%) by Trained Channel with CNN, training on single-channel, test on single-channel dataset \cite{Shin2017Open}, mental arithmetic task} \footnotesize
\scalebox{0.9}{
\begin{tabular}{p{0.15\columnwidth}|p{0.21\columnwidth}|p{0.07\columnwidth}|p{0.21\columnwidth}|p{0.07\columnwidth}}
\toprule
\textbf{Trained Channel }& \textbf{Top predicted\newline Channel} & \textbf{Top\newline ACC} & \textbf{Low predicted\newline Channel} & \textbf{Low ACC} \\ \midrule
F7 &  F3 & 90.69  & P3 &  75.00      \\
AFF5h &  POO2 & 71.03  & F3 & 62.41 \\
F3 &  FCC5h, FCC4h & 80.00  & AFF5h & 74.13    \\
AFp1 & FCC4h  & 75.86  & Cz & 65.69         \\ 
AFp2 & PPO1h  &  73.27 & F7, P7 &  66.20        \\ 
AFF6h &  FCC4h, CCP6h & 71.72  & CCP4h & 65.17      \\ 
F4 &  FCC5h & 83.27  & Cz &   74.65        \\ 
F8 & CCP3h  &  82.24 & Pz &   75.69        \\ 
AFF1h & P3  & 65.86  & Pz &   60.00         \\ 
AFF2h &  FCC4h & 71.03  & CCP6h &  65.00        \\ 
Cz &  FCC6h & 80.00  & P3 &  72.41        \\       
Pz &  FCC6h & 90.86  & F8 &  86.20          \\ 
FCC5h &  PPO2h & 85.51  & AFF2h &  76.55         \\ 
FCC3h & F7  & 84.48  & Cz &  75.17         \\ 
CCP5h &  POO2 &  78.62 & AFF1h &  72.58          \\ 
CCP3h & FCC6h  & 77.58  & P3 &  68.62          \\ 
T7 &  FCC6h &  90.34 & F8 &  86.37          \\ 
P7 & FCC6h  &  87.93 & P3 &  81.55          \\ 
P3&  AFp2 &  86.72 & FCC3h &  81.37          \\ 
PPO1h &  FCC6h & 88.96  & P3 &  84.65         \\ 
POO1 & CCP3h, PPO2h  & 87.93  & Cz &  82.75         \\ 
POO2 &  FCC6h & 88.27  & Cz &  83.96         \\ 
PPO2h &  FCC6h &  90.34 & Cz &   84.82        \\ 
P4 & P8  & 86.37  & AFF2h &    81.03       \\ 
FCC4h & AFp2  &  83.79 & Cz &   76.89        \\ 
FCC6h &  PPO2h & 86.37  & F3 &  80.17         \\ 
CCP4h & FCC6h  & 77.24  & P3 &   68.96        \\ 
CCP6h &  FCC4h & 78.27  & AFF5h &   71.37        \\ 
P8 &  T7, FCC4h &  88.27 & PPO1h &  82.41         \\
T8 &  AFp2 &  91.55 & AFF2h & 85.86          \\ 
\bottomrule
\end{tabular}
}
\label{tab:channel_performance_shin2017open_ma_approach2}
\end{table}

\begin{table}[htbp]
\setlength{\tabcolsep}{5pt}
\centering
\caption{Top and Low Test Channel Performance (median \%) by Trained Channel, dataset \cite{Shin2017Open}, motor imagery task}
\label{tab:channel_performance_shin2017open_mi_approach2}
\scalebox{0.9}{
\begin{tabular}{p{0.15\columnwidth}|p{0.21\columnwidth}|p{0.07\columnwidth}|p{0.21\columnwidth}|p{0.07\columnwidth}}
\toprule
\textbf{Trained Channel}& \textbf{Top predicted\newline Channel} & \textbf{Top\newline ACC} & \textbf{Low predicted\newline Channel} & \textbf{Low ACC} \\ \midrule
F7 &  AFp1, P3 & 62.59  & CCP4h &  56.72      \\
AFF5h& F3  & 59.31  & CCP6h &  53.45\\
F3& P8  &  65.86 & POO1 &  58.10   \\
AFp1& AFF1h  & 62.76  & P3 &  55.86       \\ 
AFp2& F4  & 60.86  & FCC4h & 55.69         \\ 
AFF6h& F3  & 59.66  & Cz &  54.31     \\ 
F4& Pz  & 64.31  & FCC4h &  58.28         \\ 
F8& P8  & 63.79  & CCP4h &   57.93       \\ 
AFF1h& AFF1h, CCP3h  &  54.14 & PPO1h & 49.31           \\ 
AFF2h&  P8 &  58.45 & T7 & 53.10         \\ 
Cz&  F3 & 64.83  & AFF1h & 57.93        \\       
Pz& FCC4h  & 73.45  & CCP6h &  65.86          \\ 
FCC5h& F3  & 65.17  & CCP6h &  59.48         \\ 
FCC3h& POO2  & 65.52  & POO1, CCP6h &  58.79         \\ 
CCP5h& AFp2  & 63.10  & PPO1h & 57.24          \\ 
CCP3h& F3  & 63.45  & T7 &  56.72          \\ 
T7& F7  & 72.76  & CCP6h &   65.86         \\ 
P7&  T7 & 71.37  & AFF6h &  64.13          \\ 
P3& POO1  & 70.34  & AFF5h &   63.79         \\ 
PPO1h& AFF2h  & 70.00  & CCP6h &  64.31         \\ 
POO1& AFF1h  & 68.96  & PPO2h, AFF5h & 63.10          \\ 
POO2& F3  & 67.93  & PPO2h &  61.20         \\ 
PPO2h& AFF2h  & 71.37  & P7, P3 & 65.17          \\ 
P4& FCC5h, CCP4h  &  70.00 & AFF5h &   64.48        \\ 
FCC4h& F3  &  65.17 & CCP6h & 59.48          \\ 
FCC6h& FCC5h  &  66.72 & AFF2h  &  61.55        \\ 
CCP4h& F4, P8  & 60.17  & Cz, F8 &  56.03         \\ 
CCP6h& FCC5h  & 63.62  & F8 &   57.07        \\ 
P8& AFF1h  &  69.65 & CCP6h  &  63.62         \\
T8& FCC3h  &  72.41 & AFF5h &  68.45         \\ 
\bottomrule
\end{tabular}
}
\end{table}

By analyzing the performance of single-channel EEG across different cognitive tasks, we observed task-dependent variation in channel efficiency. Specifically, channels located in frontal and fronto-central regions, such as F4, AFF1h and AFF2h (as reported in the \cite{shin2018eyes} dataset) and FCC3h and FCC4h (as found in the \cite{Shin2017Open} dataset), demonstrated high classification accuracies of 98.33\% and 97.93\%, respectively for the two datasets in mental arithmetic. This high performance can be attributed to their involvement in executive functions and working memory, which are key components of mathematical problem solving. Similarly, central and parietal regions, including Pz, PPO2h, CCP3h and CCP4h, performed better in motor imagery tasks, as reported in the \cite{Shin2017Open} dataset, achieving the highest accuracy of 96.55\%. This finding is consistent with their known role in sensorimotor processing.

Conversely, channels located in temporal and parietal areas, such as P7, T7 and T8, were consistently among the least accurate for mental arithmetic tasks in both \cite{shin2018eyes} and \cite{Shin2017Open} datasets. This could indicate a lower engagement of these areas in the cognitive aspects of mental arithmetic processing. For motor imagery tasks, channels in temporal regions, such as TP7, T7 and T8, reported in the \cite{ma2020multi} dataset, also showed poor performance, highlighting their minimal involvement in the neural representation of motor functions. However, parietal region, such as P4 and P6 proved more significant in classifying motor imagery tasks within the \cite{ma2020multi} dataset, with the highest accuracy reaching 96\%. These results not only deepen our understanding of the neurophysiological mechanisms of mental arithmetic and motor imagery, but also have important implications for optimizing EEG channel selection in BCI applications, suggesting that a personalized approach to channel selection, taking into account task-specific cortical engagement, could improve system performance and efficiency.

Another benefit of correctly classifying at the EEG channel level is to identify the regions of interest (ROI). Accurately identifying the brain regions is crucial for obtaining adequate signals for a BCI application. By delineating the exact region of interest, it is possible to optimize the electrode placement on the scalp to detect brain activities and subsequently reduce the need for heavy EEG headsets on patients. Therefore, reducing the number of electrodes by selecting the best classification performance from electrodes is essential for controlling the BCI system with a limited number of channels \cite{Hong2018Feature}.

\subsection{Single channel training, single channel testing}

In addition to the multi-channel training approach, this study also adopted a second methodology aimed at refining channel-specific classifications. In this approach, the CNN model is trained on data from a single channel, allowing it to specialize and learn in depth the unique patterns of that specific channel. This training method is particularly beneficial for detecting and highlighting channel-specific phenomena that might be ignored when training on multiple channels. Moreover, by focusing on a single channel, the model becomes highly specialized in identifying patterns crucial for specific tasks, thus improving its effectiveness for these applications. After training, the model is then tested on all the other channels, using its specialized knowledge to generalize and apply the learned patterns to classify cognitive tasks accurately.
This method complements our first approach by offering precise insights into how each individual channel can discriminate between different brain states. Furthermore, by training and testing the model in this manner that is both focused and extensive - first focusing on a single channel, then generalizing to all the others - we enhance the model's adaptability. The model not only learns the specific features of a particular channel but also tests its ability to apply this knowledge to a diverse set of channels, encountering a wide range of data variations. 

The results of the channel-level classification using the CNN, where each model was trained on a single channel and tested on all others, are detailed in Tables 
(\ref{tab:perf_channel_approach2_shin2018}, \ref{tab:channel_performance_shin2017open_ma_approach2}, \ref{tab:channel_performance_shin2017open_mi_approach2}).  
The results presented in these tables show variability in classification performance between different EEG channels for the two cognitive tasks. This suggests that some EEG channels hold more discriminative information than others for the mental arithmetic and motor imagery tasks.

Table \ref{tab:perf_channel_approach2_shin2018} demonstrates that channels P7, AFF6h, AFp1, F8, T7, P3, F3, AFF2h, C3, P4, AFp2, F4, and AFF1h achieved a classification accuracy of 100\% when they were trained on channels POO1, POO2, P4, P8, C4, and T8. Similarly, Table \ref{tab:channel_performance_shin2017open_ma_approach2} indicates that channels AFp1, FCC6h, and F3 reached the highest accuracy of 91.55\% after being trained on data from F7, Pz, T7, T8, and PPO2h. The highest performance of AFp1, FCC6h, and F3 is consistent with their EEG locations, which are situated over frontal regions engaged during mental arithmetic tasks. These results suggest that although the training was performed on specific channels, the CNN model was capable to be well generalized to other channels for the mental arithmetic task.

The results of the top predicted and low predicted test channel performance for the motor imagery task, based on the trained channel are show in Table \ref{tab:channel_performance_shin2017open_mi_approach2}. the table reveals that channels such as FCC4h, F7, AFF2h, and FCC3h, when trained on Pz, T7, PPO1h, and PPO2h, yielded the highest accuracies, indicating their strong neural correlation with motor imagery task. In contrast, channels including PPO1h and T7 showed lower performance, suggesting that not all regions provide the valuable information for this task. These findings highlight the need for strategic channel selection in BCI systems to improve the detection of neural activity related to motor imagery.

Our study investigates the feasibility of using single-channel EEG data for cognitive task classification to improve wearable BCI technologies. We found that single-channel EEG systems can decode complex neurological phenomena, offering significant advantages for real-world applications due to their ease of installation and portability. 

\section{Conclusion}

In this study, we investigated the performance of a shallow CNN model using frequency and temporal data representation to classify mental arithmetic and motor imagery tasks using single-channel EEG data at the trial level. Our results highlight the effectiveness of our approach not only improve classification accuracy and reduce the number of electrodes necessary for a BCI but also to reduce computational load. 
We applied our methods to three distinct datasets on two cognitive tasks - mental arithmetic and motor imagery. The findings demonstrate that single-channel approaches can effectively handle EEG data across varied experimental paradigms with different recording properties and a diverse range of EEG channel settings, suggesting significant promise for future BCI and brain-monitoring applications.

\bibliographystyle{ieeetr} 
\bibliography{bib}

\end{document}